\documentclass[conference,letterpaper,final]{IEEEtran}
\IEEEoverridecommandlockouts

\usepackage[utf8]{inputenc}
\usepackage[T1]{fontenc}
\usepackage[english]{babel}
\usepackage{microtype}
\usepackage{blindtext}
\usepackage{csquotes}

\usepackage[style=ieee,sorting=none,backend=biber,citestyle=numeric-comp]{biblatex}
\usepackage[pdftex]{graphicx}
\graphicspath{{../../fig/}}

\usepackage[acronym]{glossaries}
 
\newacronym{fpga}{FPGA}{field programmable gate array}
\newacronym{bss}{\mbox{BSS}}{\mbox{BrainScaleS}}
\newacronym{bss2}{\mbox{BSS-2}}{\mbox{BrainScaleS-2}}
\newacronym{bss1}{\mbox{BSS-1}}{\mbox{BrainScaleS-1}}
\newacronym{pcb}{PCB}{printed circuit board}
\newacronym{snn}{SNN}{spiking neural network}
\newacronym{soc}{SoC}{system-on-chip}
\newacronym{asic}{ASIC}{application-specific integrated circuit}
\newacronym{ram}{RAM}{random-access memory}
\newacronym{trl}{TRL}{technology readiness level}
\newacronym{cpu}{CPU}{central processing unit}
\newacronym{simd}{SIMD}{single instruction multiple data}
\newacronym{mgt}{MGT}{multi-gigabit transceiver}
\newacronym{i2c}{$\text{I}^\text{2}\text{C}$}{inter-integrated circuit}
\newacronym{jtag}{JTAG}{jtag}
\newacronym{lvds}{LVDS}{lvds}
 
\usepackage[obeyFinal]{todonotes}
\usepackage[commandnameprefix=always]{changes}

\usepackage{ragged2e}

\usepackage{ifthen}
\makeatletter
\newcounter{IEEE@bibentries}
\renewcommand\IEEEtriggeratref[1]{\renewbibmacro{finentry}{\stepcounter{IEEE@bibentries}\ifthenelse{\equal{\value{IEEE@bibentries}}{#1}}
    {\finentry\@IEEEtriggercmd}
    {\finentry}}}
\makeatother

\usepackage{siunitx}
\PassOptionsToPackage{hyphens}{url}
\usepackage[
	hidelinks,
	pdftitle={},
	pdfauthor={},
]{hyperref}
\usepackage[capitalise]{cleveref}
\begin{document}

\title{The \acrlong{bss2} multi-chip system:\\Interconnecting continuous-time neuromorphic compute substrates\\
\thanks{This work has received funding from the EC Horizon 2020 Framework Programme under grant agreement Nos. 720270, 785907 and 945539  (HBP), the EC Horizon Europe Framework Programme under grant agreement 101147319 (EBRAINS 2.0), and the Deutsche Forschungsgemeinschaft (DFG, German Research Foundation) under Germany’s Excellence Strategy EX 2181/1-390900948 (the Heidelberg STRUCTURES Excellence Cluster).
}
}

\author{\IEEEauthorblockN{Joscha Ilmberger and Johannes Schemmel}
\IEEEauthorblockA{\textit{Kirchhoff-Institute for Physics} and \textit{Institute of Computer Engineering} \\
\textit{Heidelberg University}\\
Heidelberg, Germany \\
joscha.ilmberger@kip.uni-heidelberg.de, johannes.schemmel@ziti.uni-heidelberg.de}
}

\maketitle

\begin{abstract}
	The \acrlong{bss2} SoC integrates analog neuron and synapse circuits with digital periphery, including two \acrshortpl{cpu} with \acrshort{simd} extensions.
Each ASIC is connected to a Node-FPGA, providing experiment control and Ethernet connectivity.
This work details the scaling of the compute substrate through \acrshort{fpga}-based interconnection via an additional Aggregator unit.
The Aggregator provides up to \num{12} transceiver links to a backplane of Node-FPGAs, as well as \num{4} transceiver lanes for further extension.
Two such interconnected backplanes are integrated into a standard \SI{19}{in} rack case with \SI{4}{U} height together with an Ethernet switch, system controller and power supplies.
For all spike rates, chip-to-chip latencies---consisting of four hops across three \acrshortpl{fpga}---below \SI{1.3}{\micro\second} are achieved within each backplane.
 \end{abstract}


\section{Introduction}
\label{sec:introduction}

\begin{figure}[!th]
	\vspace{-2mm}
	\center{\includegraphics[width=\columnwidth]{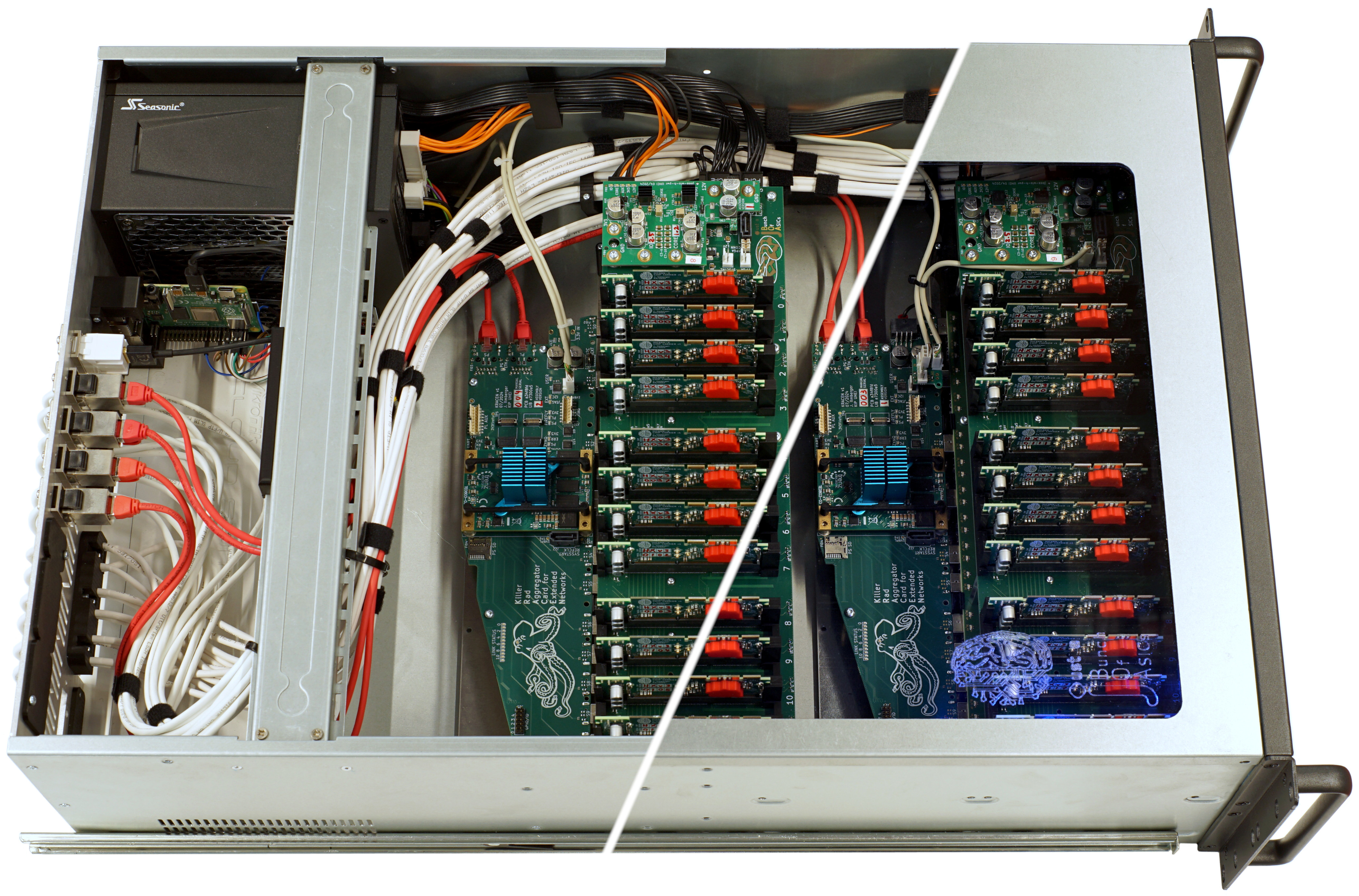}
		\\
		\vspace*{2mm}
		\includegraphics[width=0.6\columnwidth]{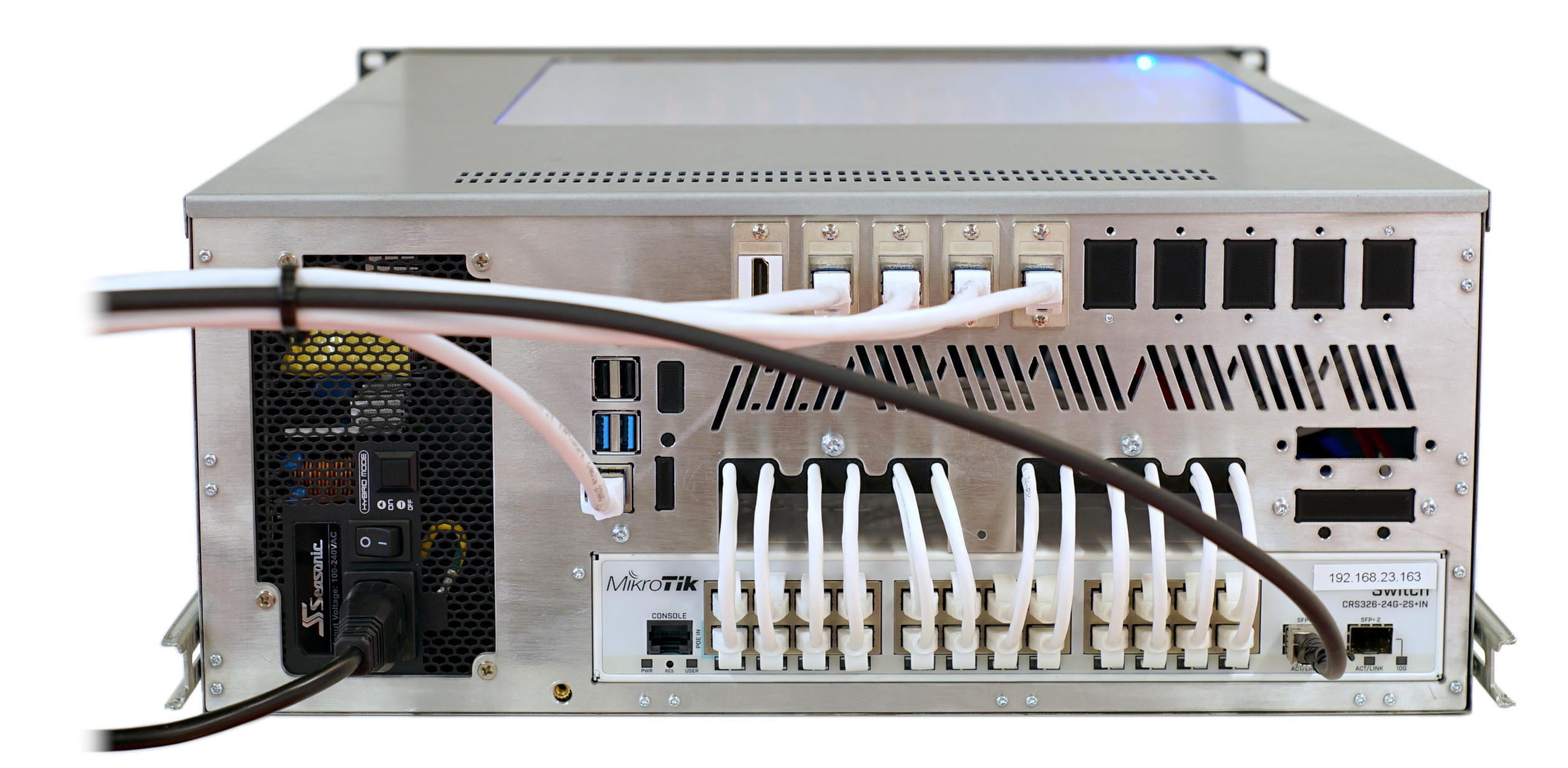}
		\hfill
		\includegraphics[width=0.28\columnwidth]{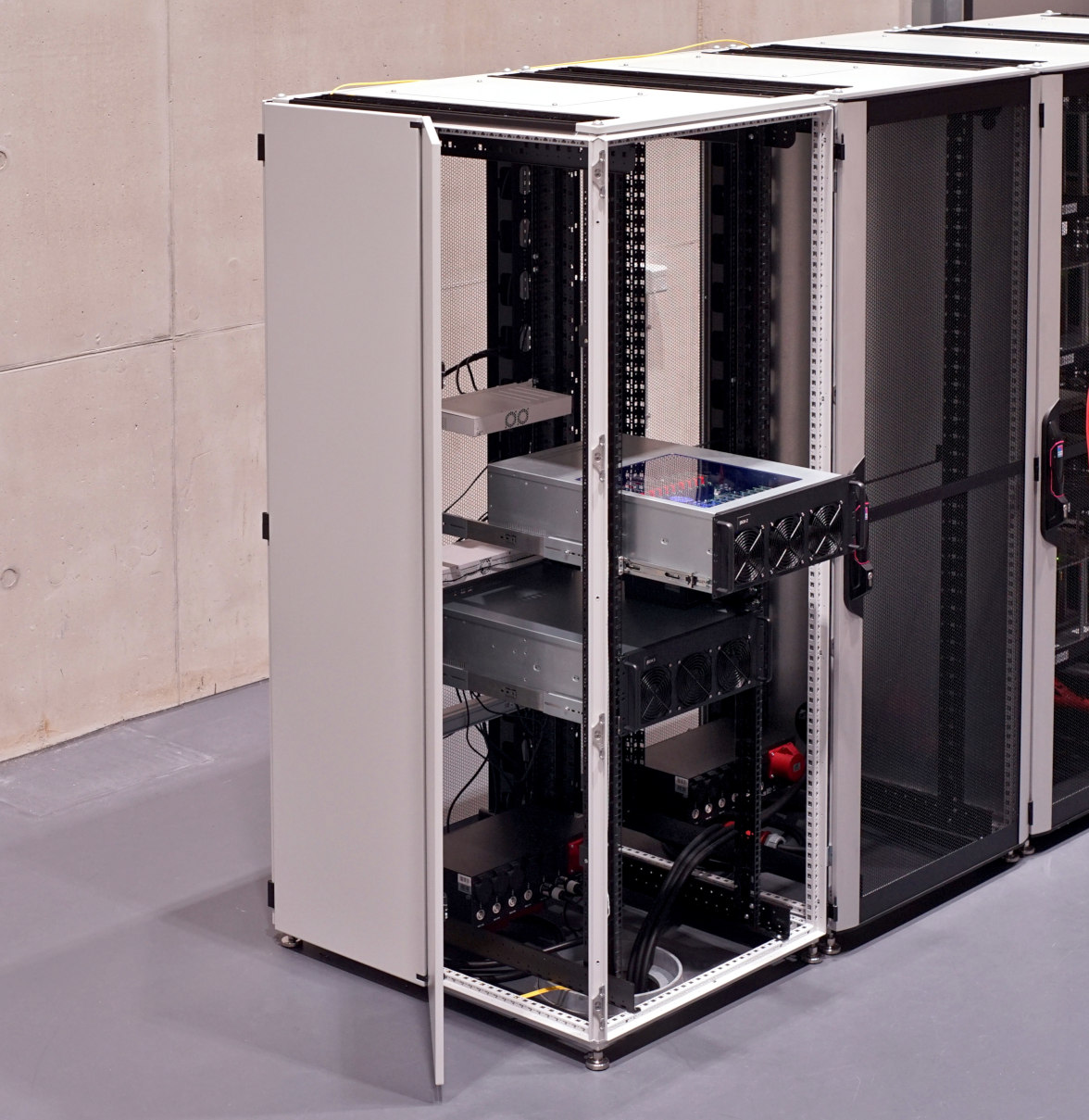}
		\hspace*{5mm}
		\caption{\textit{Top:} The \acrlong{bss2} multi-chip system.
		Fully equipped, it will consist of two backplanes with \num{12} interconnected \acrlong{bss2} \acrshortpl{soc} each, corresponding to a total of \num{12} thousand neurons and \num{3} million synapse circuits.
		\textit{Bottom left:} Back panel view with integrated Ethernet switch, system controller and ATX power supply.
		\textit{Bottom right:} Deployment of first systems at the European Institute for Neuromorphic Computing (EINC) at Heidelberg University.
		A second-layer interconnect between all backplanes inside a rack is envisioned.
		\label{fig:rackcase}
		\vspace{-3mm}
		}
	}
\end{figure}

Neuromorphic computing aims to bridge the gap between classical compute architectures and information processing found in neurobiology.
Specifically, \glspl{snn} promise energy savings due to the sparsity of their communication, which is the bottleneck for many modern compute tasks.
Hardware systems optimized for \gls{snn} execution typically target the acceleration of specific workloads in either low-power edge applications\,\parencite{frenkel2019morphic,moradi2018dynaps,richter2024dynapse2,yao2024spike,pei2019tianjic,davies2018loihi} or a data-center compute context\,\parencite{davies2018loihi,Khan2008,gonzales2024spinnaker}.
The improvement of training algorithms and model parameter tuning methodologies remains an active area of research, especially for large-scale systems beyond the limitations of single compute substrates.

\acrfull{bss2} is a mixed-signal neuromorphic architecture targeting both application regimes\,\parencite{pehle2022brainscales2}.
The current generation of the \acrshort{bss2} \acrshort{asic} integrates \num{512} neuron and \num{131072} synapse circuits with digital periphery, including two \acrshortpl{cpu} with \acrshort{simd} extensions.
Since the dynamics of the emulated physical models run roughly \num{1000}\,-\,fold accelerated, communication and experiment control is handled by an \acrshort{fpga}.
Currently, the compute substrate is limited to chip size, which cannot be used in a resource-multiplexing fashion during runtime due to the time-continuous nature of the analog circuits.
While there are ongoing efforts to develop direct chip-to-chip interconnection\,\parencite{ilmberger2024flexible}, top-down scaling via \acrshortpl{fpga} promises a fast and flexible solution for at least \num{120} interconnected \acrshortpl{asic}, enabling the research of training methodologies for large-scale analog hardware.
Due to the acceleration factor of the architecture, the spike latency between \acrshortpl{asic} constitutes the most important system optimization target.
The following sections present the multi-chip system, spike routing architecture, and the characterization of basic operating figures.

\begin{figure*}[!ht]
	\center{\includegraphics[width=176mm]{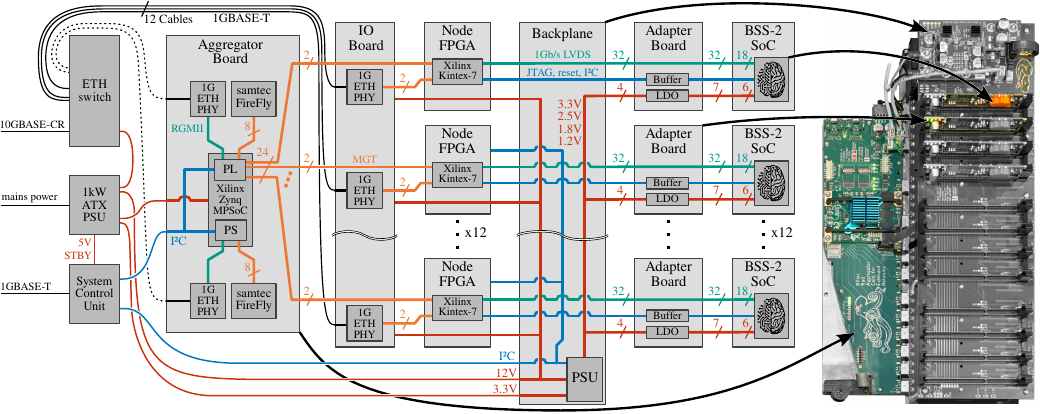}
		\caption{Overview of the neuromorphic multi-chip system.
		One backplane connects up to \num{12} \acrlong{bss2} SoCs with one Node-\acrshort{fpga} each via adapter boards\textsuperscript{\ref{footnote:tud}}.
		The adapter board contains all necessary periphery of the neuromorphic SoC such as level shifters, LDO regulators and DACs.
		Network sizes beyond a single chip can be achieved by interconnecting the transceivers of all Node-\acrshortpl{fpga} to an additional Aggregator unit.
		The star topology allows for symmetric delays below \SI{1.3}{\micro\second} of any source neuron to off-chip target synapse.
		Together with off-the-shelf components such as an ATX power supply, Ethernet switch, and ARM-based system controller, up to two backplanes can be combined into an air-cooled \SI{4}{U} high \SI{19}{in} rack case.
		This unit requires only mains power and Ethernet uplinks to operate.
		\label{fig:system_overview}
		}
	}
\end{figure*}

\section{System description}
\label{sec:system}

\Cref{fig:rackcase} provides an overview of the presented system.
It features a standard \SI{19}{in} air-cooled rack case and requires only mains power and Ethernet uplinks to operate, allowing comfortable data-center-like operation.
For ease of maintenance, all cables are routed through an energy chain, allowing access during operation via full-extension slides.

More details on data and power interconnection inside the system are provided in \Cref{fig:system_overview}.
The system-control unit is responsible for powering up all components, basic configuration and continuous monitoring of their state.
Hardware access is split across multiple resettable \acrshort{i2c} chains for increased robustness.
In contrast, all neuromorphic experiment control is facilitated by the Node-\acrshort{fpga} via \acrshort{jtag} and a custom source-synchronous \acrshort{lvds} high-speed interface\footnote{\label{footnote:tud}The Node-\acrshort{fpga} board, IO-Board (see \cref{fig:system_overview}), custom high-speed interface and PLL of the \acrshort{bss2} \acrshort{asic} were developed at the Chair of Highly-Parallel VLSI Systems and Neuro-Microelectronics of the Dresden University of Technology.} towards the \acrshort{bss2} \acrshort{asic}.
The current \acrshort{asic} makes use of \num{18} of the \num{32} available high-speed signals, allowing future system upgrades with more than one chip per adapter board.
To support this significant power draw increase of the chip carriers, all LDO regulators can be bridged directly to power rails distributed across the backplane.
The adapter board features configurable LDO voltages, power monitoring and multiple DAC channels, which are controlled by the Node-\acrshort{fpga} via \acrshort{i2c}.

Multi-\acrshort{fpga} operation of \acrshort{bss2} requires synchronization of clocks, as well as the experiment real-time section starting point.
For this purpose, a common \SI{50}{\mega\hertz} reference clock and an additional system start signal are routed symmetrically from the Aggregator to all Node-\acrshortpl{fpga} via the IO-Board.

The Node-\acrshortpl{fpga} and IO-Boards were initially developed and manufactured for the \acrlong{bss1} wafer-scale system and are re-used in this presented multi-chip architecture.
This cost optimization dictates the form factor, as well as the power budget of the system with the \num{24} Node-\acrshortpl{fpga} requiring roughly \SI{400}{\watt} to operate.
In total, the system is comprised of up to \num{80} instances of \num{9} unique \acrshort{pcb} types, \num{40} data cables and \num{10} power cables, depending on the individual configuration.
 \section{Routing implementation}
\label{sec:routing}

The \acrlong{bss} architectures distinguish different spike communication layers.
Layer-1 handles single spikes in a real-time fashion with minimum buffer sizes and thus incurs loss in case of continued congestion.
In contrast, layer-2 communication allows for packing of up to three spikes for bandwidth efficiency, larger buffer sizes and uses a tagged system time for jitter compensation.
Finally, layer-3 is used for non-real-time connectivity using classical packaging and networking methods.

Communication between the \gls{bss2} \acrshort{asic} and Node-\acrshort{fpga} is implemented in a layer-2 fashion, while on-chip spike traffic follows the layer-1 approach.
Since the presented multi-chip extension features deterministic delays by design, it can omit timestamping to fully utilize the available bandwidth of a single transceiver lane.
This approach is contrary to previous work, which focused on interconnection at higher layer levels\,\parencite{thommes2022demonstrating}.
The transceiver latency is optimized significantly by choosing 8b10b encoding at its highest allowed line rate of \SI{5}{\giga\bit\per\second}, rather than the overall highest possible bandwidth of \SI{8}{\giga\bit\per\second} using 64b66b encoding.

\begin{figure*}
	\center{\includegraphics{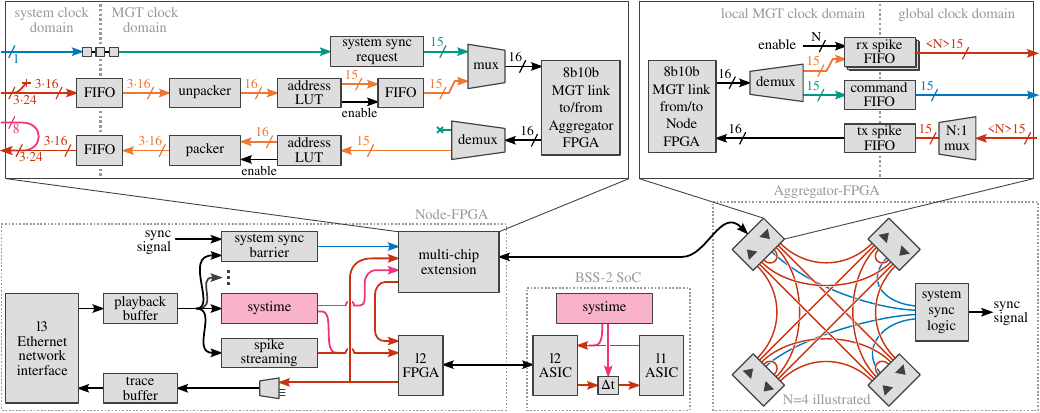}
		\caption{Multi-chip extension to the existing Node-\acrshort{fpga} design and routing logic.
		Just before the experiment real-time section, a system synchronization barrier command gets executed from the playback buffer, resulting in a request being sent via the \acrfull{mgt} link to the aggregator.
		Once the aggregator has received the request from all participating Node-\acrshortpl{fpga}, an external synchronization signal is toggled, causing the playback execution to continue.
		During the real-time section, output spikes of the \acrshort{bss2} \acrshort{soc} neurons coming out of the layer-1 crossbar get a timestamp attached and are transported the Node-\acrshort{fpga} via the layer-2 link.
		The multi-chip extension listens in on this traffic, discards the timestamp, unpacks the spikes and uses a Block-\acrshort{ram} based lookup for \SI{15}{\bit} labels and routing enable.
		Inside the aggregator, spikes are broadcasted in an all-to-all connectivity scheme with static enables for each route.
		Spikes can be sent and received each clock cycle with the exception of clock-compensation pauses by the transceiver.
		All spikes that are sent back to a Node-\acrshort{fpga} pass a reverse lookup to \SI{16}{\bit} \acrshort{bss2} \acrshort{asic} spike labels and are packed.
		After attaching the lower eight bit of the current system time, which is synchronized with the \acrshort{asic}, the spikes can be sent through the layer-2 link.
		With a transceiver user clock of \SI{250}{\mega\hertz}, the maximum theoretical spike throughput of the \acrshort{bss2} \acrshort{asic} can be sustained.
		The implementation of the routing logic is the simplest possible and should be seen as a baseline for testing more complex schemes.
		\label{fig:k7node_gt_adapter}
		}
	}
\end{figure*}

\Cref{fig:k7node_gt_adapter} details how the existing Node-\acrshort{fpga} design is extended for multi-chip operation, as well as the implemented routing scheme.
The \acrshort{fpga} module currently deployed on the aggregator unit provides only four transceiver lanes, thus interconnecting four chips.
Pin-compatible modules with \num{16} transceiver lanes are available, allowing for a future plug-and-play upgrade.
Benchmark tests have shown that the all-to-all connectivity could be scaled to the maximum of \num{12} Node-\acrshort{fpga} plus \num{4} extension lanes.
Nevertheless, more complex and resource-efficient scalable routing schemes can be evaluated on the presented platform.

All experiments are timed using the system clock domain based on the globally distributed reference clock to avoid drifts of the timebase within the system.
Using the additional system start signal (see \cref{sec:system}), the starting point of the real-time section can be synchronized to within one system clock cycle of \SI{8}{\nano\second}.
While every participating Node-\acrshort{fpga} notifies the Aggregator of their readiness using a command message via the \gls{mgt} link, the response is distributed using this external signal.
The system synchronization logic inside the Aggregator features configurable timeout- and refractory periods as fault recovery mechanisms.
This approach is decentralized and symmetric, as no Node-\acrshort{fpga} requires a different configuration or plays a different role in the synchronization process.

All output spikes of the \acrshort{bss2} neuron circuits are routed through a layer-1 crossbar, being sent to on-chip synapses or back to the Node-\acrshort{fpga} via the layer-2 link to be stored in the trace buffer for the experiment user.
This data stream consisting of up to three parallel events with \SI{16}{\bit} labels and \SI{8}{\bit} timestamps is tapped by the multi-chip extension.
While the timestamp could be used to compensate parts of the link jitter it is currently omitted and the parallel data stream is passed to the faster \SI{250}{\mega\hertz} \gls{mgt} clock domain.
Here, all units operate on single events, matching the maximum sustained spike rate of the \acrshort{bss2} \acrshort{asic} link.
The \gls{mgt} link accepts \SI{16}{\bit} per clock cycle with the exception of clock-compensation pauses.
To allow command messages and future protocol extensions, only \SI{15}{\bit} are available for real-time event traffic.
Since the \acrshort{bss2} spike labels consist of \SI{16}{\bit} and not all traffic should be routed off-chip, some mapping is required.
For maximum flexibility, derived from \parencite{stradmann2024closing}, a full \SI{16}{\bit} to \SI{16}{\bit} lookup is implemented via Block-\acrshortpl{ram}, of which one bit is interpreted as routing enable.
Finally, all enabled labels are passed on to the Aggregator via the \gls{mgt} link.
The Aggregator strips off command messages and implements all-to-all connectivity with configurable enables per route.
This architecture is flexible enough to map non-recurrent multi-layer networks where every \acrshort{bss2} chip encompasses few layers.
If more demanding use-cases arise, the routing logic can easily be expanded by some mapping between spike labels and route enables.
In the reverse direction towards another Node-\acrshort{fpga}, all received spike labels are re-mapped in a full \SI{15}{\bit} to \SI{17}{\bit} lookup, again, including one enable bit.
The remaining \SI{16}{\bit} are now interpreted as \acrshort{bss2} spike labels and are passed on to the system clock domain after packing for bandwidth-efficiency.
With an attached timestamp, these events are merged with the user-defined spike streaming from the playback memory and sent to the \acrshort{bss2} \acrshort{asic} via the layer-2 link.
Here, a small buffer and comparison with an expected link delay offers limited jitter compensation before the spikes are passed on to the layer-1 crossbar.

 \section{Measurements}
\label{sec:measurements}

\begin{figure}
	\center{\includegraphics[width=\columnwidth]{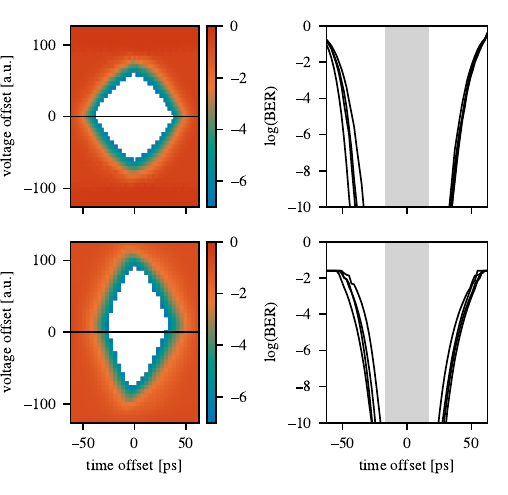}
		\caption{Data eye and bathtub analysis of the Node-\acrshort{fpga} to Aggregator (\textit{top}) and reverse direction (\textit{bottom}) multi-gigabit transceiver link using the manufacturer's analysis tool.
		The tested \SI{8}{\giga\bit\per\second} data rate is the maximum supported by the Node-\acrshort{fpga}.
		The data eye is exemplary for one link, whereas the bathtub curves are shown for all links in addition to the recommended margin.
		Bit-error rate tests up to $10^{-15}$ were successfully executed, suggesting a smaller required margin.
		The shown transceiver configuration was not optimized for power efficiency.
		\label{fig:kracen_ibert}
		}
	}
\end{figure}

The \gls{mgt} links between Node-\acrshortpl{fpga} and Aggregator span \num{4} \acrshortpl{pcb} and consequently \num{3} connectors with a total substrate length between \SI{250}{\milli\meter} and \SI{375}{\milli\meter}.
Therefore, the signal integrity is verified using the tools provided by the manufacturer.
\Cref{fig:kracen_ibert} shows measurements at the maximum line rate with no detected errors for more than one day of continuous runtime.
Since the links are used at a much lower \SI{5}{\giga\bit\per\second} line rate and 8b10b encoding to minimize latency, the error margin is increased even further.
This allows spike data---in contrast to control data---to be transferred across the \gls{mgt} link without the overhead of error-checking codes or methods.

\begin{figure}
	\center{\includegraphics[width=68mm]{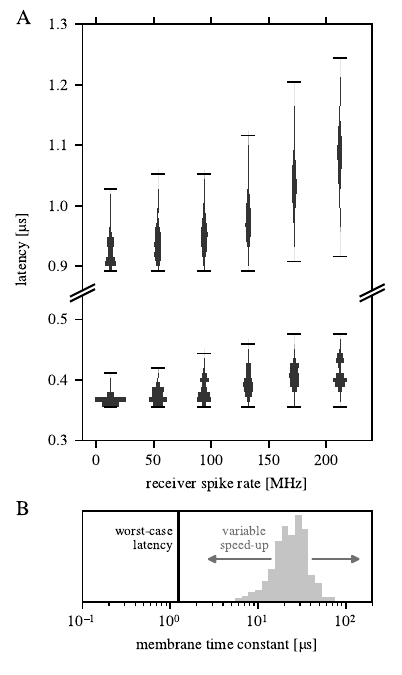}
		\hfill
		\includegraphics[width=17mm]{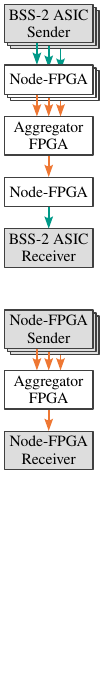}
		\caption{\textbf{(A)} Measurement of the latency of $2^{15}$ spikes with a 3:1 fan-in between Node-\acrshortpl{fpga} (\textit{bottom}) and between \acrshort{bss2} \acrshortpl{asic} (\textit{top}) for a range of regular rates up to congestion of the receiver.
		In the worst regime, the total event jitter constitutes roughly \SI{15}{\percent} of the median delay.
		The visible discretization of the distributions corresponds to the \SI{8}{\nano\second} \acrshort{fpga} system clock period used to measure the latency.
		\textbf{(B)} With the default hardware speed-up of $10^{3}$, the routing latency is one order of magnitude below typical membrane time constants measured in biology\,\parencite{sunkin2012allen}.
The speed-up factor can be chosen within certain bounds due to large circuit parameter calibration ranges\,\parencite{billaudelle2022accurate}, reducing or increasing the model parameters with respect to the fixed routing latency as indicated.
		\label{fig:kracen_latency}
		}
	}
\end{figure}

The total spike latency is characterized using a range of regular rates with three senders and one receiver on the four-chip prototype setup (see \cref{fig:system_overview}, right side).
\Cref{fig:kracen_latency} shows the resulting distributions with bandwidth-independent, deterministic delays of the \acrshort{fpga}-based interconnect, increasing only by a handful of system clock cycles due to congestion at the multiplexer inside the Aggregator.
The two \gls{mgt} link hops take \SI{0.3}{\micro\second}, with the rest of the inter-\acrshort{fpga} delay distributed across the mapping and routing logic inside the sender, receiver and Aggregator.
Roughly \SI{60}{\percent} of this additional delay is caused by counter synchronizations at clock domain crossings, with the rest added by the packing logic, pipeline stages of the address LUT and multiplexer arbitration.
The additional spike round-trip-time to the \acrshort{asic} cannot be optimized significantly in this work.
While the lower tail of the distribution could be squashed to some extent using similar jitter compensation techniques as deployed in the layer-2 to layer-1 boundary of the chip, this would however increase the latency overall and not compensate the upper tail.
The on-chip jitter compensation can be seen in the histogram below \SI{100}{\mega\hertz} spike rates and ceases to have any notable effect above due to link congestion.

 \section{Discussion and Outlook}
\label{sec:discussion}

This work presents a system designed to scale an accelerated analog \gls{snn} architecture beyond individual substrates.
The approach of routing spikes via \acrshortpl{fpga} is intended as a fast and flexible development path, enabling the evaluation of routing architectures and training algorithms at scale.
The insights gained can be transferred to future systems with direct \acrshort{asic}-interconnection\,\parencite{ilmberger2024flexible}, increasing density and power-efficiency.
With this goal in mind, the multi-chip system is designed in an adaptable fashion, allowing the connection of whole mesh-networks of \acrshortpl{asic} to one adapter board and Node-\acrshort{fpga}, again, interconnected by the presented Aggregator unit.

With the current generation of \acrshort{bss2} \acrshortpl{asic}, at least \num{120} chips can be interconnected by a single second-layer node, combining \num{10} Aggregator units within one rack in a star topology.
This results in more than \num{61} thousand neurons and \num{15} million synapses with an expected chip-to-chip latency increase of roughly \SI{0.4}{\micro\second}, due to two additional transceiver hops.
The system size and density numbers are largely the result of historical design choices and re-use of existing hardware components, rather than fundamental architectural constraints.
To the best of our knowledge, the presented system will constitute the second-largest analog continuous-time \acrshort{snn} system, surpassed only by the \acrlong{bss1} wafer-scale system\,\parencite{schemmel2010iscas}.

With the default speed-up factor of \num{1000} of the \acrshort{bss2} architecture in mind, the presented latency between \SI{0.9}{\micro\second} and \SI{1.3}{\micro\second} is roughly one order of magnitude below common membrane time constants found in biology\,\parencite{sunkin2012allen,tripathy2014neuroelectro}.
It has to be noted that the speed-up factor is not a fixed number and can be shifted down within certain hardware parameter ranges in case the spike latency proves to be an issue for certain networks or models, as demonstrated by the example of the membrane time constant in \Cref{fig:kracen_latency}B.
Non-recurrent feed-forward networks with whole layers---or parts thereof---mapped to individual chips may be able to circumvent the routing latency entirely.

Finally, we note that the presented system is not optimized for energy efficiency, as the Node-\acrshortpl{fpga} introduce a power overhead of more than one order of magnitude relative to the neuromorphic \acrshort{asic}.
 
\section*{Contributions}
Joscha Ilmberger conceptualized and developed the presented system, performed the measurements and wrote the manuscript.
Johannes Schemmel is the principal architect of BrainScaleS-2 and gave conceptual advice.
 
\section*{Acknowledgment}
\label{sec:acknowledgments}

The authors wish to thank all present and former members of the Electronic Visions research group contributing to the BrainScaleS-2 neuromorphic platform, specifically Yannik Stradmann for fruitful discussions and Julian G\"oltz for early system testing.
We especially thank Andreas Gr\"ubl, Lars Sterzenbach, Burak Ayhan, Nikolas Merklinger, Jan Niklas Schneider and Alexander Dobler of the Electronics Workshop of the Kirchhoff-Institute for Physics, Christian Herdt, David Jansen and Julia Bing of the Mechanics Workshop of the Kirchhoff-Institute for Physics, as well as Markus Dorn and Ralf Achenbach of the ASIC laboratory of Heidelberg University.
 
\IEEEtriggeratref{14}
\printbibliography

\end{document}